\documentclass[aps,prl,nofootinbib,preprint]{revtex4} 
\usepackage{graphicx}

\begin{document}

\title{Ginzburg -- Landau Expansion in BCS -- BEC Crossover Region of
Disordered Attractive Hubbard Model}

\author{$^1$E.Z. Kuchinskii, $^1$N.A. Kuleeva, $^{1,2}$M.V. Sadovskii}


\affiliation{$^1$Institute for Electrophysics, Russian Academy of Sciences, Ural Branch,\\ 
Amundsen str. 106, Ekaterinburg 620016, Russia\\
$^2$M.N. Mikheev Institute for Metal Physics, Russian Academy of Sciences, Ural Branch,\\ 
S. Kovalevky str. 18, Ekaterinburg 620290, Russia}


\begin{abstract}

We have studied disorder effects on the coefficients of Ginzburg -- Landau (GL)
expansion for attractive Hubbard model within the generalized 
DMFT+$\Sigma$ approximation for the wide region of the values of attractive 
potential $U$ --- from the weak-coupling limit, where superconductivity is
described by BCS model, towards the strong coupling, where superconducting
transition is related to Bose -- Einstein condensation (BEC) of compact Cooper
pairs.

For the case of semi-elliptic initial density of states disorder influence on 
the coefficients $A$ and $B$ before the square and the fourth power of the
order parameter is universal for at all values of electronic correlations and
is related only to the widening of the initial conduction band (density of
states) by disorder. Similar universal behavior is valid for superconducting 
critical temperature $T_c$ (the generalized Anderson theorem) and specific heat 
discontinuity at the transition.

This universality is absent for the coefficient $C$ before the gradient term,
which in accordance with the standard theory of ``dirty'' superconductors is
strongly suppressed by disorder in the weak-coupling region, but can slightly 
grow in BCS -- BEC crossover region, becoming almost independent of disorder in 
the strong coupling region. This leads to rather weak disorder dependence of the
penetration depth and coherence length, as well as the slope of the upper
critical magnetic field at $T_c$, in BCS -- BEC crossover and strong coupling
regions. 

PACS:71.10.Fd, 74.20.-z, 74.20.Mn

\end{abstract}

\maketitle


\section{Introduction}

Ilya Mikhailovich Lifshitz was one of the creators of the modern theory of
disordered systems \cite{IML}. Among his numerous contributions in this field we
only mention the general formulation of the concept of self -- averaging
\cite{IML_UFN} and the method of optimal fluctuation for the description of
``Lifshitz tails'' in the electron density of states \cite{IML_JETP}. These ideas
and approaches are widely used now in many fields of the theory of disordered
systems, even those which initially were outside the scope of his personal
scientific interests. 

The studies of disorder effects in superconductors have a rather long history.
The pioneer works by Abrikosov and Gor'kov \cite{AG_impr,AG_imp,AG_mimp} and
Anderson \cite{And_th} had been devoted to the limit of weakly disordered
metal ($p_Fl\gg 1$, where $p_F$ is Fermi momentum and $l$ is the mean free path)
and weakly coupled superconductors, well described by BCS theory \cite{Genn}.
The notorious Anderson theorem \cite{And_th,Genn} on $T_c$ of superconductors
with ``normal'' (spin independent) disorder was proved in this limit under the
assumption of self -- averaging superconducting order -- parameter
\cite{Genn,SCLoc_PRep,SCLoc}. The generalizations for the case of strong enough
disorder ($p_Fl\sim 1$) were also mainly done under the same assumption, though
it can be explicitly shown, that self -- averaging of the order parameter is
violated close to Anderson metal -- insulator transition \cite{SCLoc_PRep,SCLoc}.
Here, the ideas originating from Ref. \cite{IML_JETP} are of primary importance
\cite{SadBul87}.

The problem of superconductivity in disordered systems in the limit of strongly
coupled Cooper pairs, including the region of BCS -- BEC crossover, was not well
studied until recently. In fact, the problem of superconductivity in the case of
strong enough pairing interactions was considered for a long enough time
\cite{Leggett}. Significant progress here was achieved by Nozieres and
Schmitt-Rink \cite{NS}, who proposed an effective method to study the crossover
from BCS behavior in the weak coupling region towards Bose -- Einstein
condensation of Cooper pairs in the strong coupling region. One of the simplest
models, where we can study the BCS -- BEC crossover, is Hubbard model with
attractive interaction. The most successful theoretical approach to describe
strong electronic correlations in Hubbard model (both repulsive and attractive)
is the dynamical mean field theory (DMFT) \cite{pruschke,georges96,Vollh10}. The
attractive Hubbard model was already studied within this approach in a number of
papers \cite{Keller01,Toschi04,Bauer09,Koga11,JETP14}. However, there are only
few papers, where disorder effects in BCS -- BEC crossover region were taken
into account.

In recent years we have developed the generalized DMFT+$\Sigma$ approach to
Hubbard model \cite{JTL05,PRB05,FNT06,UFN12}, which is very convenient for the
studies of different ``external'' (with respect to DMFT) interactions, such as
pseudogap fluctuations \cite{JTL05,PRB05,FNT06}, disorder scattering
\cite{HubDis,HubDis2} and electron -- phonon interaction \cite{e_ph_DMFT}).
This approach is also well suited to the analysis of two -- particle properties,
such as dynamic (optical) conductivity \cite{HubDis,PRB07}.
In Ref. \cite{JETP14} we have used this approach to analyze the single --
particle properties of the normal (non -- superconducting) phase and optical
conductivity of the attractive Hubbard model. Further on, DMFT+$\Sigma$
approach was used to study disorder influence on superconducting transition
temperature, which was calculated within Nozieres -- Schmitt-Rink approach
\cite{JTL14,JETP15}. 

The general review of DMFT+$\Sigma$ approach was given in Ref. \cite{UFN12},
and the review of this approach to disordered Hubbard model (both repulsive
and attractive) was recently presented in Ref. \cite{JETP16_rev}.

In this paper we investigate Ginzburg -- Landau (GL) expansion for disordered
attractive Hubbard model including the BCS -- BEC crossover region and the limit
of strong coupling. Coefficients of GL -- expansion in BCS -- BEC crossover
region were studied in a number of papers \cite{Micnas01,Zwerger92,Zwerger97},
but there were no previous studies of disorder effects, except our recent
paper \cite{JETP16}, where we have considered only the case of homogeneous GL --
expansion and demonstrated certain universal behavior of GL --
coefficients on disorder (reflecting the generalized Anderson theorem).
Below we mainly concentrate on the study of the GL -- coefficient
before the gradient term, where such universal behavior is just absent.
Here we limit ourselves to the case of weak enough disorder ($p_Fl\gg 1$),
neglecting the effects of Anderson localization, which can significantly change
the behavior of this coefficient in the limit of strong disorder
\cite{SCLoc_PRep,SCLoc}.

\section{Hubbard model within DMFT+$\Sigma$ approach}

We shall consider the disordered paramagnetic Hubbard model with attractive
interaction. The Hamiltonian is written as:
\begin{equation}
H=-t\sum_{\langle ij\rangle \sigma }a_{i\sigma }^{\dagger }a_{j\sigma
}+\sum_{i\sigma }\epsilon _{i}n_{i\sigma }-U\sum_{i}n_{i\uparrow
}n_{i\downarrow },  
\label{And_Hubb}
\end{equation}
where $t>0$ is the transfer integral between the nearest neighbors on the
lattice, $U$ is the Hubbard -- like on site attraction,
$n_{i\sigma }=a_{i\sigma }^{\dagger }a_{i\sigma }^{{%
\phantom{\dagger}}}$ is electron number operator at site $i$, $a_{i\sigma }$ 
($a_{i\sigma }^{\dagger}$) is electron annihilation (creation) operator at
$i$-th site and spin $\sigma$. Local energy levels $\epsilon _{i}$ are assumed to be 
independent and random at different sites.  To use the standard ``impurity''
diagram technique we assume the Gaussian statistics for energy levels $\epsilon _{i}$:
\begin{equation}
\mathcal{P}(\epsilon _{i})=\frac{1}{\sqrt{2\pi}\Delta}\exp\left(
-\frac{\epsilon_{i}^2}{2\Delta^2}
\right)
\label{Gauss}
\end{equation}
Parameter $\Delta$ here is the measure of disorder strength, while the Gaussian
random field of energy levels introduces the ``impurity'' scattering, which is
considered using the standard approach, using the averaged Green's functions
\cite{Diagr}.

The generalized DMFT+$\Sigma$ approach \cite{JTL05,PRB05,FNT06,UFN12} adds to
the standard DMFT \cite{pruschke,georges96,Vollh10} an additional ``external''
electron self -- energy $\Sigma_{\bf p}(\varepsilon)$ (in general case momentum
dependent), which is produced by additional interactions outside the DMFT,
which gives an effective procedure to calculate both single -- particle and
two -- particle properties \cite{HubDis,PRB07,JETP16_rev}. The success of this
approach is related to choice of the single -- particle Green's function in the
following form:
\begin{equation}
G(\varepsilon,{\bf p})=\frac{1}{\varepsilon+\mu-\varepsilon({\bf p})-\Sigma(\varepsilon)
-\Sigma_{\bf p}(\varepsilon)},
\label{Gk}
\end{equation}
where $\varepsilon({\bf p})$ -- is the ``bare'' electron dispersion, while the
full self -- energy is the additive sum the local self -- energy
$\Sigma (\varepsilon)$, determined from DMFT, and ``external''
$\Sigma_{\bf p}(\varepsilon)$. Thus we neglect all the interference processes
between of Hubbard and ``external'' interactions. This allows us to conserve the
general structure of self -- consistent equations of the standard DMFT
\cite{pruschke,georges96,Vollh10}. At the same time, at each step of DMFT
iterations the ``external'' self -- energy $\Sigma_{\bf p}(\varepsilon)$ is
recalculated using some approximate calculation scheme, corresponding to the
form of additional interaction, while the local Green's function is dressed by
$\Sigma_{\bf p}(\varepsilon)$ at each step of DMFT procedure.

Here, in the impure Hubbard model, the ``external'' self -- energy entering
DMFT+$\Sigma$ is taken in the simplest form (self -- consistent Born
approximation), which neglects all diagrams with intersecting lines of impurity
scattering, so that:
\begin{equation}
\Sigma_{\bf p}(\varepsilon)\to\tilde\Sigma(\varepsilon)=
\Delta^2\sum_{\bf p}G(\varepsilon,{\bf p}),
\label{BornSigma}
\end{equation}
where $G(\varepsilon,{\bf p})$ is the single -- electron Green's function
(\ref{Gk}) and $\Delta$ is the amplitude of site disorder.

To solve the effective Anderson impurity model of DMFT throughout this paper
we used the numerical renormalization group (NRG) algorithm \cite{NRGrev}.
All calculations below were done for the case of the quarter -- filled band
(n=0.5 electrons per lattice site).

Further on we shall consider the model of the ``bare'' conduction band with
semi -- elliptic density of states (per unit cell and single spin projection):
\begin{equation}
N_0(\varepsilon)=\frac{2}{\pi D^2}\sqrt{D^2-\varepsilon^2}
\label{DOSd3}
\end{equation}
where $D$ defines the band half -- width. This is a rather good approximation for
three -- dimensional case.

In Ref. \cite{JETP15} we have given an analytic proof that in DMFT+$\Sigma$
approximation for disordered Hubbard model with semi -- elliptic density of
states all disorder effects in single -- particle properties, calculated
in DMFT+$\Sigma$ (with the use of self -- consistent Born approximation
(\ref{BornSigma})) are reduced to conduction band -- widening by disorder, i.e.
to the replacement (in the density of states) $D\to D_{eff}$, where $D_{eff}$
is the effective half -- width of the band  in the presence of disorder scattering:
\begin{equation}
D_{eff}=D\sqrt{1+4\frac{\Delta^2}{D^2}}.
\label{Deff}
\end{equation}
so that the ``bare'' density of states (in the absence of correlations,$U=0$)
becomes:
\begin{equation}
\tilde N_{0}(\xi)=\frac{2}{\pi D_{eff}^2}\sqrt{D_{eff}^2-\varepsilon^2}
\label{tildeDOS}
\end{equation}
conserving its semi -- elliptic form. It should be noted, that for different
models of the ``bare'' conduction band disorder can also change the form of the
density of states, so that such universal disorder effects in single -- properties
is absent. However, in the limit of strong enough disorder almost any initial
density of states actually acquires semi -- elliptic form, restoring this
universal dependence on disorder \cite{JETP15}.

The temperature of superconducting transition in attractive Hubbard model within
DMFT was calculated in a number of papers \cite{Keller01,Toschi04,Koga11},
analyzing both from the Cooper instability of the normal phase \cite{Keller01}
(divergence of Cooper susceptibility) and from the disappearance of
superconducting order parameter \cite{Toschi04,Koga11}. In Ref. \cite{JETP14} we
determined the critical temperature from instability of the normal phase
(instability of DMFT iteration procedure). The results obtained were in good
agreement with the results of Refs. \cite{Keller01,Toschi04,Koga11}. Besides
that, in Ref. \cite{JETP14} to calculate $T_c$ we have used the Nozieres --
Schmitt-Rink approach \cite{NS}, showing that this approach allows qualitatively,
though approximately, describes the BCS -- BEC crossover region. In Refs.
\cite{JTL14,JETP15} we used the combination of Nozieres -- Schmitt-Rink approach
and DMFT+$\Sigma$ for detailed studies of disorder influence on the temperature
of superconducting transition and the number of local pairs.
In this approach we determine $T_c$ from the following equation \cite{JETP15}:
\begin{equation}
1=\frac{U}{2}\int_{-\infty}^{\infty}d\varepsilon \tilde N_0(\varepsilon)
\frac{th\frac{\varepsilon -\mu}{2T_c}}{\varepsilon -\mu}.
\label{BCS}
\end{equation}
with chemical potential $\mu$ for different $U$ and $\Delta$ being determined from
DMFT+$\Sigma$ calculations, i.e. from the standard equation for the number of
electrons (band filling), defined by the Green's function (\ref{Gk}). This
allows us to find  $T_c$  for the wide range of the model parameters, including
the BCS -- BEC crossover region and the limit of strong coupling, as well as for
the different disorder levels. This reflects the physical meaning of Nozieres --
Schmitt-Rink approximation: in the weak coupling region transition temperature
is controlled by the equation for Cooper instability (\ref{BCS}), while in the
strong coupling limit it is determined by the temperature of Bose condensation
of compact Cooper pairs, which is controlled by chemical potential.

\begin{figure}
\includegraphics[clip=true,width=0.7\textwidth]{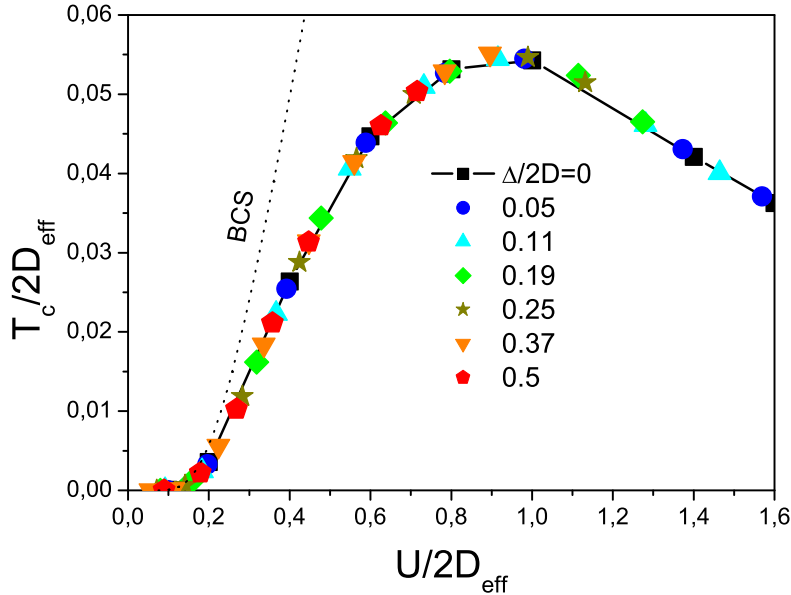}
\caption{Universal dependence of superconducting critical temperature on Hubbard
attraction for different levels of disorder.}
\label{fig1}
\end{figure}

In Fig. \ref{fig1} we show the universal dependence of superconducting critical
temperature $T_c$ on Hubbard attraction for different levels of disorder obtained
in Ref. \cite{JETP15}. This is a manifestation of the generalized Anderson theorem.
In the weak coupling region $T_c$ is well described by
BCS model (dashed line in Fig. \ref{fig1} shows $T_c$ determined by Eq. (\ref{BCS})
with chemical potential independent of $U$ and obtained for the quarter -- filled
``bare'' band), while in the strong coupling region $T_c$ is determined by the
condition of Bose condensation of Cooper pairs giving $\sim t^2/U$ dependence
(corresponding to inverse mass dependence of compact Bosons),  passing through a
characteristic maximum at $U/2D_{eff}\sim 1$ in BCS -- BEC crossover region.

\section{Ginzburg -- Landau expansion}

Ginzburg -- Landau expansion for the difference of free energies of superconducting
and normal phases can be written in the standard form:
\begin{equation}
F_{s}-F_{n}=A|\Delta_{\bf q}|^2
+q^2 C|\Delta_{\bf q}|^2+\frac{B}{2}|\Delta_{\bf q}|^4,
\label{GL}
\end{equation}
where $\Delta_{\bf q}$ is the Fourier component of the order parameter.

\begin{figure}
\includegraphics[clip=true,width=\textwidth]{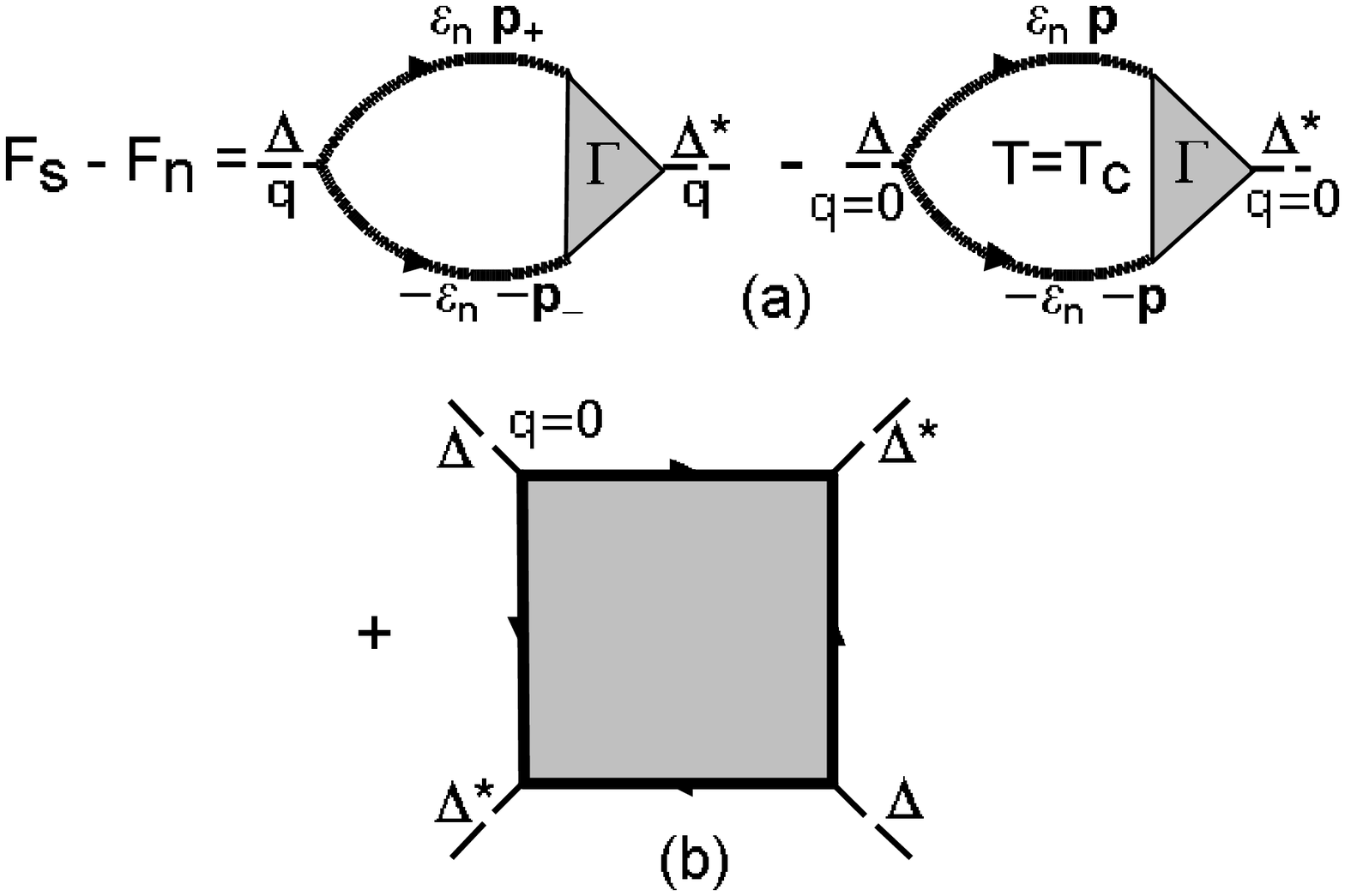}
\caption{Diagrammatic representation of Ginzburg -- Landau expansion.}
\label{diagGL}
\end{figure}

Microscopically GL -- expansion (\ref{GL}) is determined by diagrams of loop --
expansion for the free energy of electrons in an ``external'' field of random
fluctuations of order parameter with small wave -- vector ${\bf q}$
\cite{SCLoc,Diagr} shown in Fig.\ref{diagGL} (where fluctuations are
represented by dashed lines). In disordered system, the use here of the standard
impurity diagram technique implicitly assumes the self -- averaging nature of the
order parameter \cite{Genn,SCLoc_PRep,SCLoc}.
	
Within the framework of Nozieres -- Schmitt-Rink approach \cite{NS} the loops
with two and four Cooper vertices, shown in Fig.\ref{diagGL}, do not contain
contributions from attractive Hubbard interaction (as in weak coupling theory)
and are ``dressed'' only by disorder (impurity) scattering
\footnote{In the absence of disorder this approach gives the same results for
GL -- coefficients as in Refs. \cite{Micnas01,Zwerger92,Zwerger97}, where the
functional integral for free energy was analyzed via Hubbard -- Stratonovich
transformation, reducing it to the functional integral over arbitrary fluctuations
of superconducting order parameter}. However, the chemical potential here, which
has an important dependence on the strength of interaction $U$ and determines the
condition of Bose condensation of Cooper pairs, should be calculated in the
framework of DMFT+$\Sigma$ approximation, as it was done in
Refs. \cite{JTL14,JETP15} in calculations of $T_c$.

In Ref. \cite{JETP16} we have shown that in this approach GL -- coefficients
$A$ and $B$ are determined by the following expressions:
\begin{equation}
A(T)=\frac{1}{U}-
\int_{-\infty}^{\infty}d\varepsilon \tilde N_0(\varepsilon)
\frac{th\frac{\varepsilon -\mu }{2T}}{2(\varepsilon -\mu )},
\label{A_end}
\end{equation}

\begin{equation}
B=\int_{-\infty}^{\infty}\frac{d\varepsilon}{2(\varepsilon -\mu)^3}
\left(th\frac{\varepsilon -\mu}{2T}-\frac{(\varepsilon -\mu)/2T}{ch^2\frac{\varepsilon -\mu}{2T}}\right)
\tilde N_0(\varepsilon),
\label{B_end}
\end{equation}

\begin{figure}
\includegraphics[clip=true,width=0.48\textwidth]{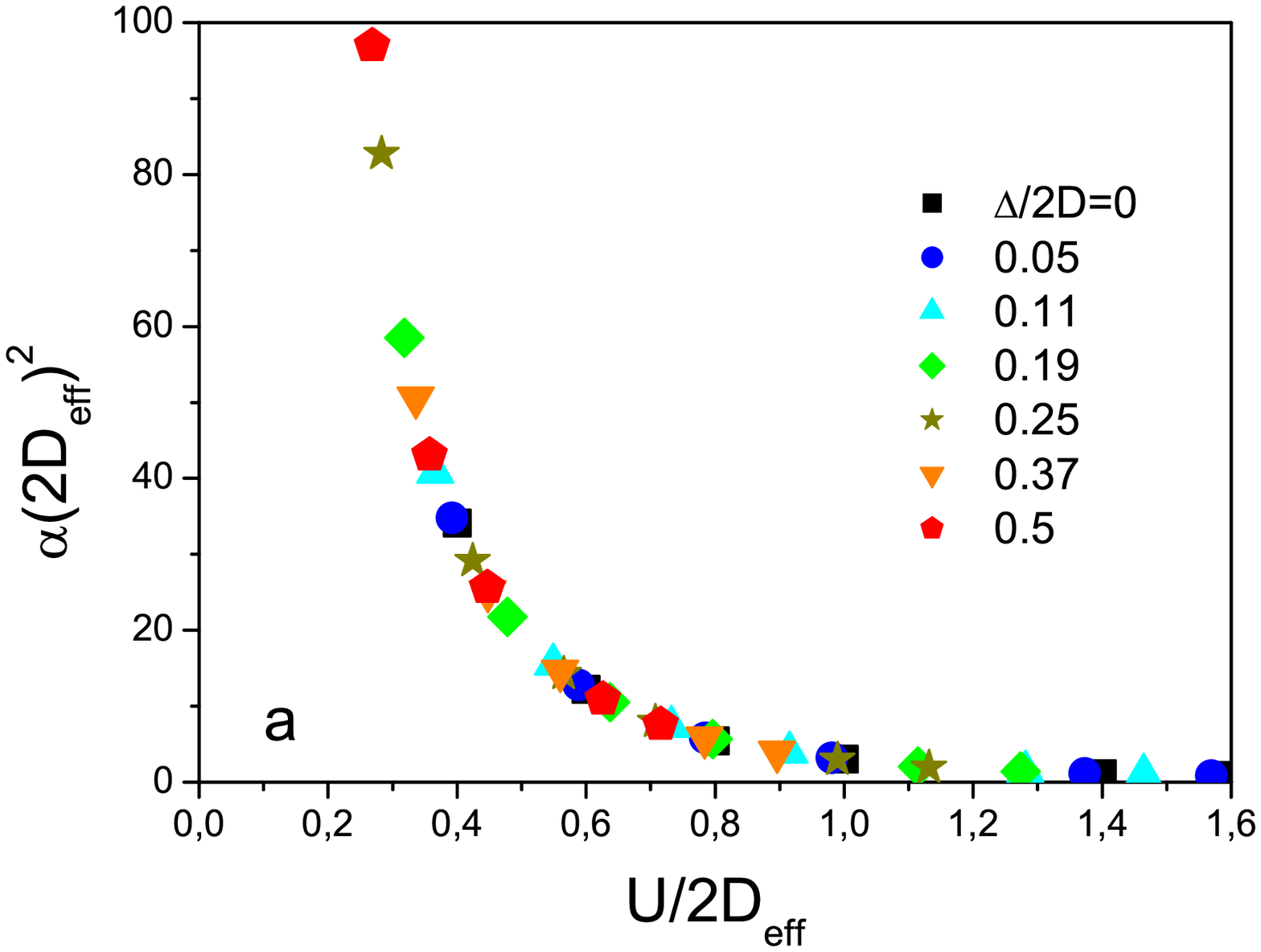}
\includegraphics[clip=true,width=0.48\textwidth]{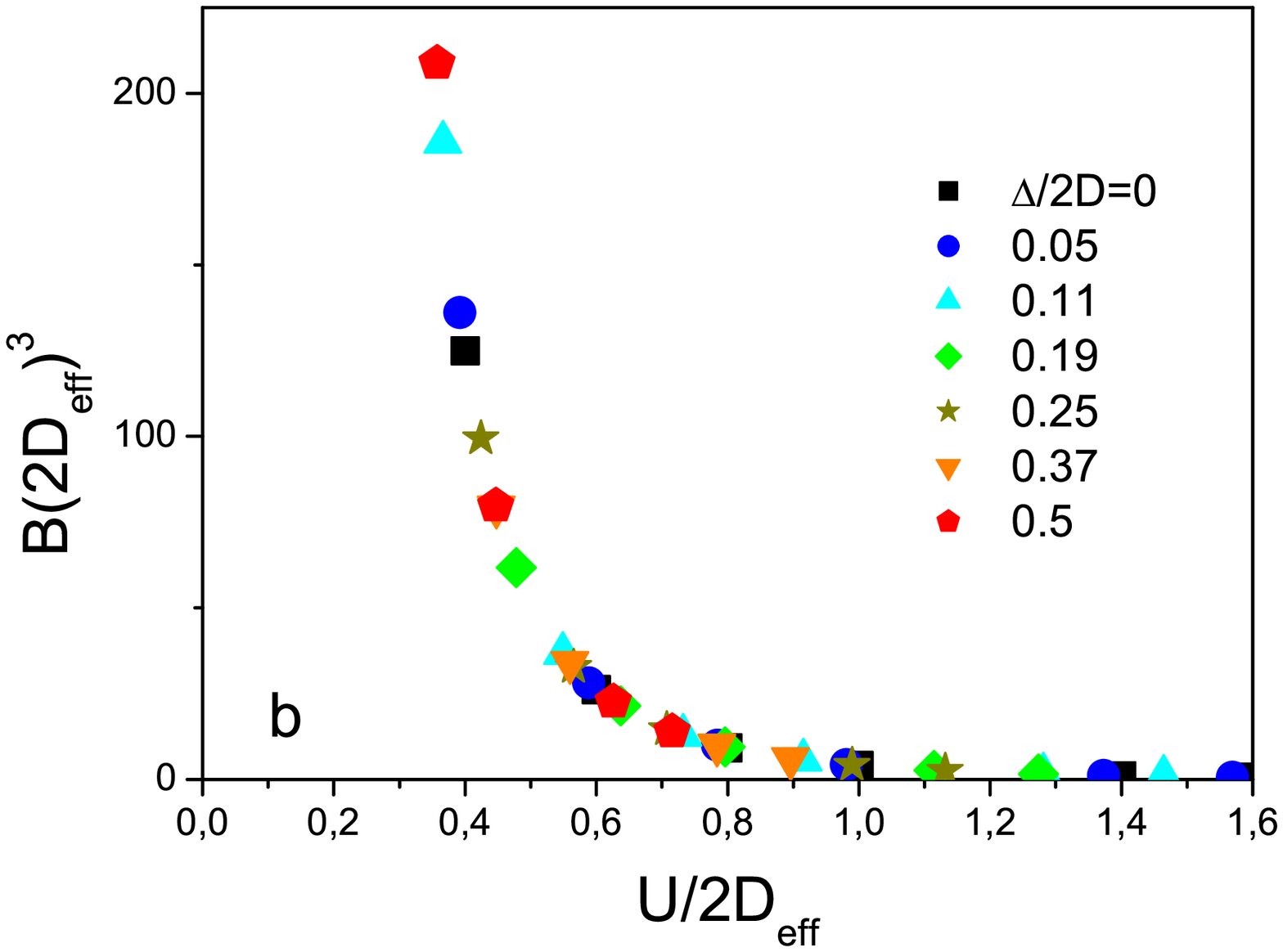}
\caption{Universal dependence of Ginzburg -- Landau coefficients $\alpha$ (a)
and $B$ (b) on the strength of Hubbard attraction for different levels of
disorder.}
\label{fig3}
\end{figure}

For $T\to T_c$ coefficient $A(T)$ takes the usual form:
\begin{equation}
A(T)\equiv \alpha(T-T_c).
\label{A2}
\end{equation} 
In BCS weak coupling limit we obtain the standard expressions for $\alpha$ and $B$ \cite{Diagr}:
\begin{equation}
\alpha_{BCS}=\frac{\tilde N_0(\mu)}{T_c},
\qquad B_{BCS}=\frac{7\zeta(3)}{8\pi^2 T_c^2}\tilde N_0(\mu).
\label{aB_BCS}
\end{equation} 
so that coefficients $A$ and $B$ are determined only by disorder widened density
of states $\tilde N_0(\varepsilon)$ and chemical potential $\mu$. Then, in the
case of semi -- elliptic density of states their dependence on disorder is
described by the simple replacement $D\to D_{eff}$  and we have universal
dependencies of $\alpha$ and $B$ (properly normalized by powers of $2D_{eff}$)
on $U/2D_{eff}$, as shown in Fig. \ref{fig3}. Both $\alpha$ and $B$ drop fast
with the growth of interaction $U/2D_{eff}$.

It should be noted that Eqs. (\ref{A_end}) and (\ref{B_end}) for coefficients
$A$ and $B$ were obtained in Ref. \cite{JETP16} using the exact Ward identities
and remain valid also in the limit of strong disorder (up to Anderson localization).
Correspondingly, in the limit of strong disorder the coefficients $A$ and $B$
depend on disorder only via appropriate dependence of the density of states.

\begin{figure}
\includegraphics[clip=true,width=0.7\textwidth]{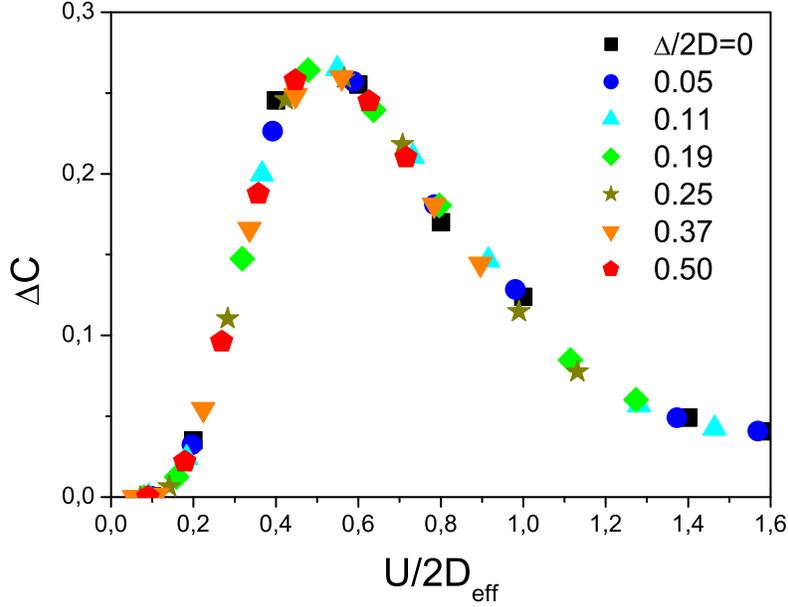}
\caption{Universal dependence of specific heat discontinuity on $U/2D_{eff}$
for different disorder levels.}
\label{fig4}
\end{figure}

Dependence on disorder, related only to the band widening by
$D\to D_{eff}$, is also observed for specific heat discontinuity at the
critical temperature \cite{JETP16}, determined by coefficients $\alpha$ and $B$:
\begin{equation}
C_s(T_c)-C_n(T_c)=T_c\frac{\alpha^2}{B}.
\label{Cs-Cn}
\end{equation}
In Fig. \ref{fig4} we show the universal dependence of specific heat discontinuity
on $U/2D_{eff}$. In BCS limit specific heat discontinuity grows with coupling,
while in BEC limit it drops with $U/2D_{eff}$, passing through maximum at
$U/2D_{eff}\approx 0.55$ in BCS -- BEC crossover region. This behavior of
specific heat discontinuity is mainly related to the similar dependence of $T_c$ 
(cf. Fig. \ref{fig1}), as $\frac{\alpha^2}{B}$ in Eq. (\ref{Cs-Cn}) only 
smoothly depends on the coupling strength.

From diagrammatic representation of GL -- expansion shown in Fig. \ref{diagGL}
it is clear, that coefficient $C$ is determined by the $q^2$ term in the
expansion of the two -- particle loop (first term in Fig. \ref{diagGL}) in
powers of $q$. Then we obtain:
\begin{equation}
C=-T\lim_{q \to 0}\sum_{n, \bf p, \bf {p'}} 
\frac{\Phi_{\bf p\bf {p'}}( \varepsilon_n,{\bf q})-\Phi_{\bf p\bf {p'}}( \varepsilon_n,0)}{q^2},
\label{C1}
\end{equation}
where $\Phi_{\bf p,\bf {p'}}( \varepsilon_n,{\bf q})$ is two -- particle Green's
function in Cooper channel ``dressed'' (in Nozieres -- Schmitt-Rink approximation)
only by impurity scattering. To determine the coefficient $C$ we again use the
exact Ward identity, derived by us in Ref. \cite{PRB07}:
\begin{equation}
G(\varepsilon_n,{\bf p_{+}})-G(-\varepsilon_n,-{\bf p_{-}})=
-\sum_{\bf p'}\Phi_{\bf pp'}
(\varepsilon_n,{\bf q})\left[(G_0^{-1}(\varepsilon_n,{\bf p'_{+}})-G_0^{-1}
(-\varepsilon_n,-{\bf p'_{-}})\right],
\label{C2}
\end{equation}
where ${\bf p_{\pm}}={\bf p} \pm \frac{\bf q}{2}$,
$G_0(\varepsilon_n,{\bf p})=\frac{1}{\varepsilon_n+\mu-\varepsilon_{\bf p}}$
is the ``bare'' single -- particle Green's function at Fermion Matsubara frequencies
$\varepsilon_n$, while $G(\varepsilon_n,{\bf p})$ is the single -- particle
Green's function ``dressed'' only by impurity scattering. Introducing the
notation $\Delta G(\varepsilon_n,{\bf p})=G(\varepsilon_n,{\bf p_+})-
G(-\varepsilon_n,-{\bf p_-})$ and using the symmetry
$\varepsilon({\bf p})=\varepsilon(-{\bf p})$ and
$G(\varepsilon_n,-{\bf p})=G(\varepsilon_n,{\bf p})$ we rewrite the Ward identity as:
\begin{equation}
\Delta G(\varepsilon_n,{\bf p})=-\sum_{\bf p'}\Phi_{\bf pp'}(\varepsilon_n,{\bf q})
(2i\varepsilon_n-\Delta \varepsilon_{\bf p'}),
\label{C3}
\end{equation}
where $\Delta \varepsilon_{\bf p}=\varepsilon_{\bf p_+}-\varepsilon_{\bf p_-}$.
Then we can perform here summation over ${\bf p}$ (also with additional
multiplication by $\Delta \varepsilon_{\bf p}$) to obtain the following system
of equations:
\begin{eqnarray}
\sum_{\bf {p}}
\Delta G(\varepsilon_n,{\bf p})&=&-2i\varepsilon_n\Phi_{0}(\varepsilon_n,{\bf q})+
\Phi_{1}(\varepsilon_n,{\bf q}) 
\nonumber\\
\sum_{\bf {p}}\Delta \varepsilon_{\bf p}\Delta G(\varepsilon_n,{\bf p})&=&
-2i\varepsilon_n\Phi_{1}(\varepsilon_n,{\bf q})
+\sum_{\bf p,p'}\Delta \varepsilon_{\bf p}\Phi_{\bf pp'}(\varepsilon_n,{\bf q})
\Delta \varepsilon_{\bf p'},
\label{C4}
\end{eqnarray}
where $\Phi_{0}(\varepsilon_n,{\bf q})=\sum_{\bf pp'}\Phi_{\bf pp'}(\varepsilon_n,{\bf q})$,
$\Phi_{1}(\varepsilon_n,{\bf q})=\sum_{\bf pp'}
\Phi_{\bf pp'}(\varepsilon_n,{\bf q})\Delta \varepsilon_{\bf p'}=
\sum_{\bf pp'}\Delta \varepsilon_{\bf p}\Phi_{\bf pp'}(\varepsilon_n,{\bf q})$.
Then, excluding $\Phi_{1}(\varepsilon_n,{\bf q})$ from this system of equations,
we obtain:
\begin{equation}
\sum_{\bf {p}}\Delta\varepsilon_{\bf p}
\Delta G(\varepsilon_n,{\bf p})=-2i\varepsilon_n\sum_{\bf {p}}
\Delta G(\varepsilon_n,{\bf p})
-(2i\varepsilon_n)^2\Phi_{0}(\varepsilon_n,{\bf q})+\sum_{\bf pp'}
\Delta\varepsilon_{\bf p}\Phi_{\bf pp'}(\varepsilon_n,{\bf q})
\Delta\varepsilon_{\bf p'}.
\label{C5}
\end{equation}
All terms in Eq. (\ref{C5}) are functions of ${\bf q}^2$. Let us write down two
lowest -- order terms of $q^2$ -- expansion of Eq. (\ref{C5}).
The $\sim q^0$ term is:
\begin{equation}
\Phi_{0}(\varepsilon_n,{\bf q}=0)=-\frac{\sum_{\bf {p}}\Delta G(\varepsilon_n,{\bf p})}
{2i\varepsilon_n}.
\label{C6}
\end{equation}
As there is no dependence on the direction of ${\bf q}$ we choose ${\bf q}=(q,0,0)$.
Then $\sim q^2$ terms are written as:
\begin{equation}
\varphi(\varepsilon_n,{\bf q}=0){(2i\varepsilon_n)}^2=\sum_{\bf {p}\bf {p'}}
v_x\Phi_{\bf {pp'}}(\varepsilon_n,{\bf q}){v_x}'-
\lim_{q \to 0} \frac
{\sum_{\bf p}\Delta\varepsilon_{\bf p}\Delta G(\varepsilon_n,{\bf p})} 
{q^2},
\label{C7}
\end{equation}
where $v_x=\frac{\partial\varepsilon_{\bf p}}{\partial p_x}$ and
$\varphi(\varepsilon_n,{\bf q}=0)=\lim_{q \to 0} \frac
{\Phi_{0}(\varepsilon_n,{\bf q})-\Phi_{0}(\varepsilon_n,0)}{q^2}$.

For weak enough disorder we can neglect localization corrections and consider
the two -- particle loop in ``ladder'' approximation for disorder scattering.
Then, due to vector nature of vertices, all vertex corrections vanish due to
angular integration and we obtain:
\begin{equation}
\sum_{\bf {p}\bf {p'}}
v_x\Phi_{\bf {pp'}}(\varepsilon_n,{\bf q}){v_x}'=
\sum_{\bf p}{v_x}^2G(\varepsilon_n,{\bf p})G(-\varepsilon_n,{\bf p})
\label{C8}
\end{equation}
For the case of isotropic spectrum we have:
\begin{equation}
\lim_{q \to 0} \frac{\sum_{\bf p}\Delta\varepsilon_{\bf p}
\Delta G(\varepsilon_n,{\bf p})}{q^2}=-\frac{1}{2}\sum_{\bf p}
\frac{{\partial}^2\varepsilon_{\bf p}}
{\partial {p_x}^2}
(G(\varepsilon_n,{\bf p})+G(-\varepsilon_n,{\bf p})).
\label{C9}
\end{equation}
As a result, we can write $C$ coefficient (\ref{C1}) as:
\begin{equation}
C=-T\sum_{n}\frac{\sum_{\bf p}{v_x}^2G(\varepsilon_n,{\bf p})G(-\varepsilon_n,{\bf p})
+\frac{1}{2}\sum_{\bf p}
\frac{{\partial}^2\varepsilon_{\bf p}}
{\partial {p_x}^2}
(G(\varepsilon_n,{\bf p})+G(-\varepsilon_n,{\bf p}))}
{{(2i\varepsilon_{n})}^2}.
\label{C10}
\end{equation}
After the standard summation over Matsubara frequencies we obtain:
\begin{equation}
C=-\frac{1}{8\pi}
\int_{-\infty}^{\infty}d\varepsilon\frac{th\frac{\varepsilon}{2T}}{\varepsilon}
\sum_{\bf p}\left({v_x}^2Im\frac {G^{R}(\varepsilon,{\bf p})G^{A}(-\varepsilon,{\bf p})}
{\varepsilon+i\delta}+\frac{{\partial}^2\varepsilon_{\bf p}}{\partial {p_x}^2}
Im\frac{G^{R}(\varepsilon,{\bf p})}{\varepsilon+i\delta}\right).
\label{C11}
\end{equation}
Finally $C$ coefficient is expressed as:
\begin{eqnarray}
C=-\frac{1}{8\pi}
\int_{-\infty}^{\infty}d\varepsilon\frac{th\frac{\varepsilon}{2T}}{\varepsilon^2}
\sum_{\bf p}
\left(
{v_x}^2Im(G^{R}(\varepsilon,{\bf p})G^{A}(-\varepsilon,{\bf p}))
+\frac{{\partial}^2\varepsilon_{\bf p}}{\partial {p_x}^2}
ImG^{R}(\varepsilon,{\bf p})
\right)+
\nonumber\\
\frac{1}{16T}\sum_{\bf p}
\left(
{v_x}^2Re(G^{R}(0,{\bf p})G^{A}(0,{\bf p}))+
\frac{{\partial}^2\varepsilon_{\bf p}}{\partial {p_x}^2}ReG^{R}(0,{\bf p})
\right).
\label{C_final}
\end{eqnarray}
The procedure to calculate velocity $v_x$ and its derivative
$\frac{{\partial}^2\varepsilon_{\bf p}}{\partial {p_x}^2}$ in the model with
semi -- elliptic density of states was discussed in detail in Ref. \cite{HubDis}.

In the absence of disorder ($\Delta=0$) we replace $G\to G_0$ and the expression
for $C$ takes the following form:
\begin{equation}
C=-\frac{1}{8\pi}
\sum_{\bf p}
\left(
\frac{{v_x}^2}{{(\varepsilon_{\bf p}-\mu)}^2}-
\frac{{\partial}^2\varepsilon_{\bf p}}{\partial {p_x}^2}
\frac{1}{\varepsilon_{\bf p}-\mu}
\right)
\left(
\frac{th\frac{\varepsilon_{\bf p}-\mu}{2T}}{\varepsilon_{\bf p}-\mu}-\frac{1}{2T}
\right)
\label{C13}
\end{equation}
In the weak coupling BCS limit in the absence of disorder the coefficient $C$
reduces to the standard expression \cite{Diagr}:
\begin{equation}
C_{BCS}=\frac{7\zeta(3)}{16\pi^2 T_c^2}N_0(\mu)\frac{v_F^2}{d},
\label{C_BCS}
\end{equation}
where $v_F$ is Fermi velocity, $d$ -- dimensionality of space. Semi -- elliptic
density of states is a good approximation for $d=3$.  As noted above disorder
influence on $C$ is not reduced to a simple replacement $N_0\to\tilde N_0$, so
that even in the BCS weak coupling limit (in contrast to coefficients $\alpha$ and
$B$ (cf. (\ref{aB_BCS})) we can not derive for $C$ a compact expression, similar
to (\ref{C_BCS}).

\section{Main results}

Let us discuss now the main results of our calculations for the gradient term
coefficient $C$ of GL -- expansion and the related physical characteristics,
such as the coherence length, penetration depth and the slope of the upper
critical magnetic field at $T_c$.

The coherence length at given temperature $\xi(T)$ determines the characteristic
scale of order -- parameter $\Delta$ inhomogeneities:
\begin{equation}
\xi^2(T)=-\frac{C}{A}.
\label{xi2}
\end{equation}
Coefficient $A$ changes its sign at the critical temperature $A=\alpha(T-T_c)$,
so that
\begin{equation}
\xi(T)=\frac{\xi}{\sqrt{1-T/T_c}},
\label{xiT}
\end{equation}
where we have introduced the coherence length as:
\begin{equation}
\xi=\sqrt{\frac{C}{\alpha T_c}}.
\label{xi0}
\end{equation}
In the weak coupling limit and in the absence of disorder it is written in the
standard form \cite{Diagr}:

\begin{equation}
\xi_{BCS}=\sqrt{\frac{C_{BCS}}{\alpha_{BCS} T_c}}=\sqrt{\frac{7\zeta(3)}
{16\pi^2 d}}\frac{v_F}{T_c}.
\label{xi_BCS}
\end{equation}

Penetration depth of magnetic field into superconductor is defined as:
\begin{equation}
\lambda^2(T)=-\frac{c^2}{32 \pi e^2}\frac{B}{A C}.
\label{lambda2}
\end{equation}
Thus:
\begin{equation}
\lambda (T)=\frac{\lambda}{\sqrt{1-T/T_c}},
\label{lambdaT}
\end{equation}
where we have introduced:
\begin{equation}
\lambda^2=\frac{c^2}{32 \pi e^2}\frac{B}{\alpha C T_c},
\label{lambda0}
\end{equation}
which in the absence of disorder has the form:
\begin{equation}
\lambda^2_{BCS}=\frac{c^2}{32 \pi e^2}\frac{B_{BCS}}{\alpha_{BCS} C_{BCS} T_c}
=\frac{c^2}{16 \pi e^2}\frac{d}{N_0(\mu)v_F^2}.
\label{lambda_BCS}
\end{equation}
Note that $\lambda_{BCS}$ does not depend on $T_c$, and correspondingly on the
coupling strength, so that it is convenient for normalization of penetration
depth $\lambda$ (\ref{lambda0}) for arbitrary $U$ and $\Delta$.

Close to $T_c$ the upper critical field $H_{c2}$ is defined via
GL -- coefficients as:
\begin{equation}
H_{c2}=\frac{\Phi_0}{2 \pi \xi^2(T)}=-\frac{\Phi_0}{2 \pi}\frac{A}{C},
\label{Hc2}
\end{equation}
where $\Phi_0=c \pi/e$ is magnetic flux quantum. Then the slope of the upper
critical field at $T_c$ is given by:
\begin{equation}
\frac{dH_{c2}}{dT}= \frac{\Phi_0}{2 \pi}\frac{\alpha}{C}.
\label{dHc2}
\end{equation}

\begin{figure}
\includegraphics[clip=true,width=0.48\textwidth]{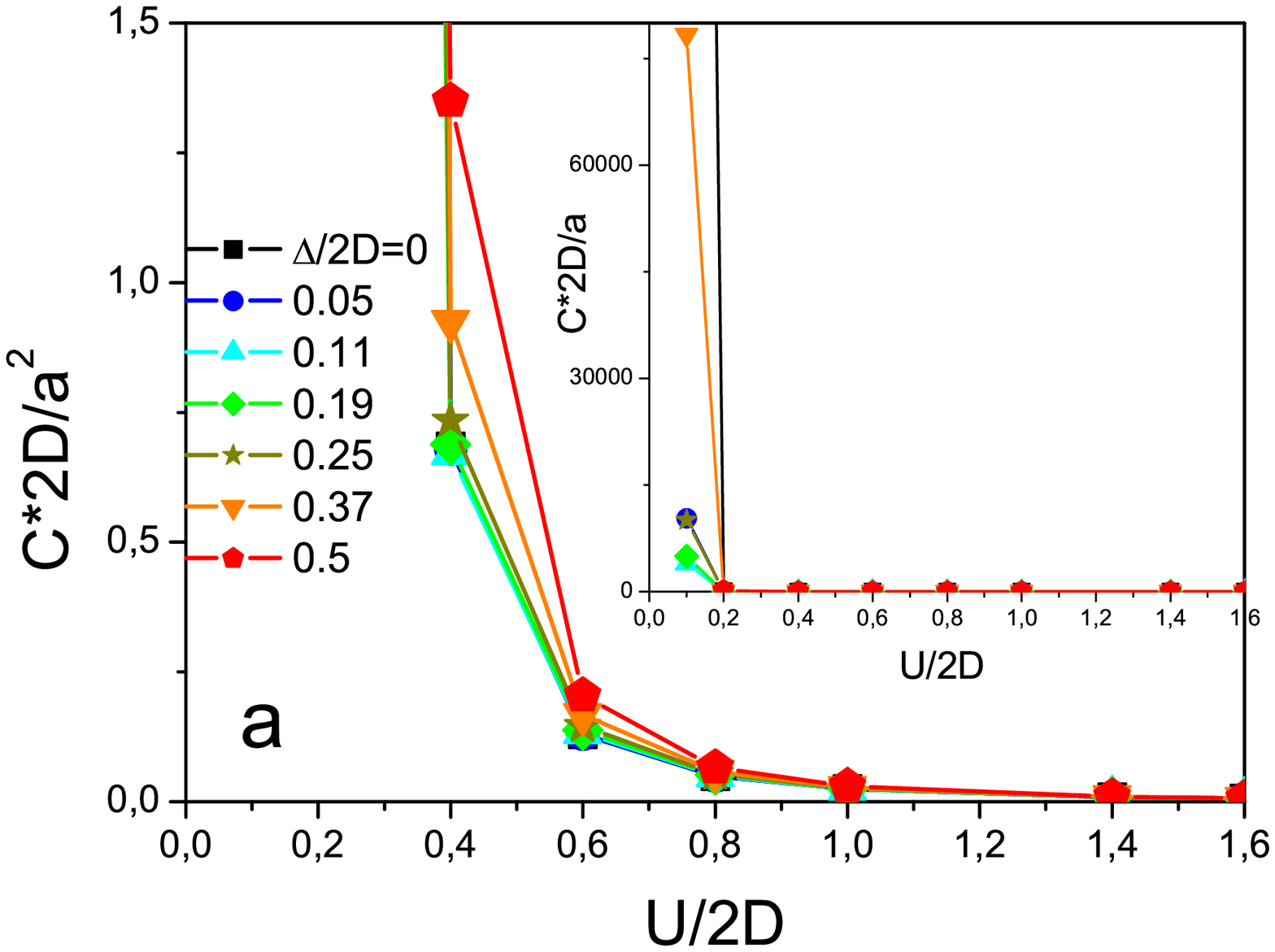}
\includegraphics[clip=true,width=0.48\textwidth]{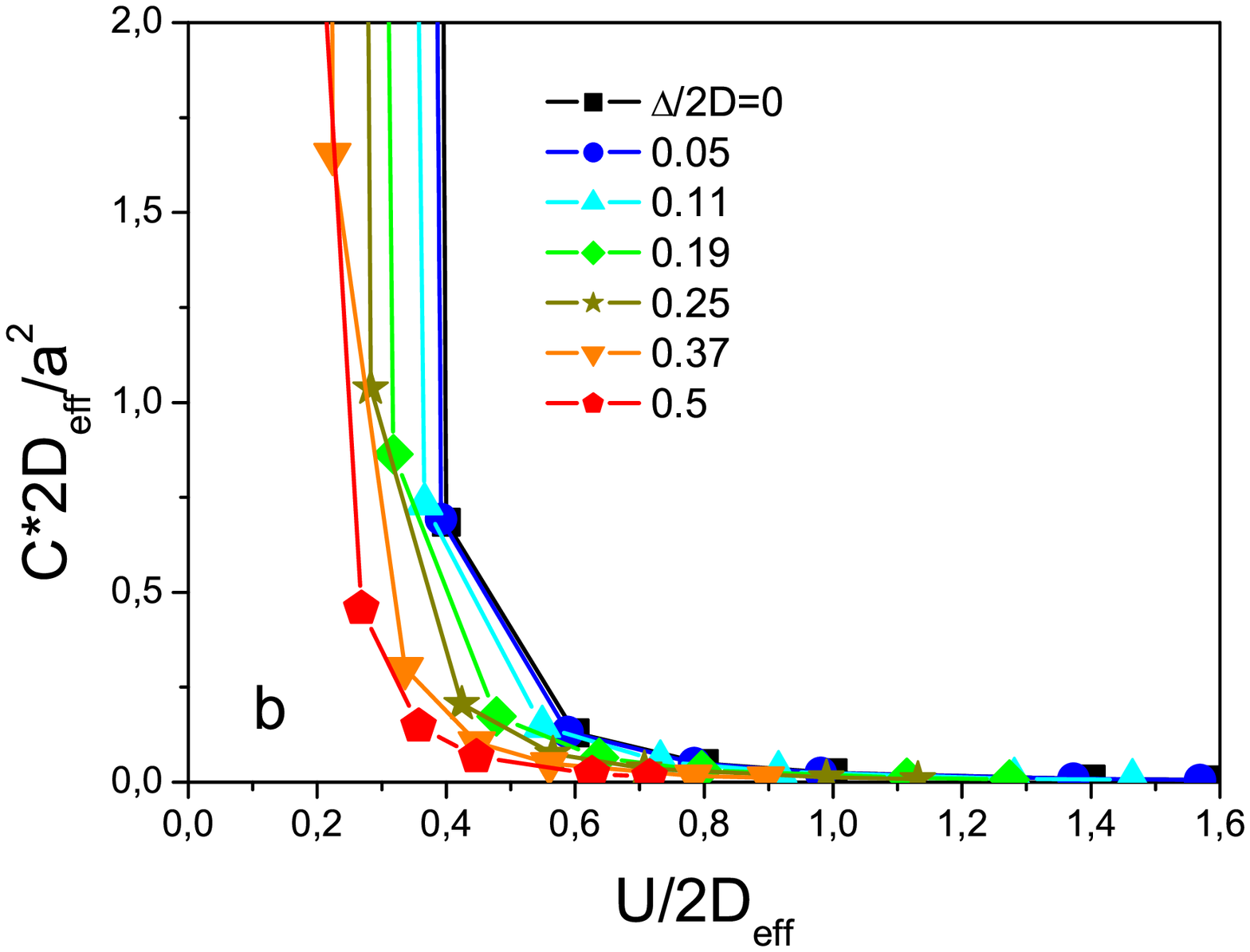}
\includegraphics[clip=true,width=0.48\textwidth]{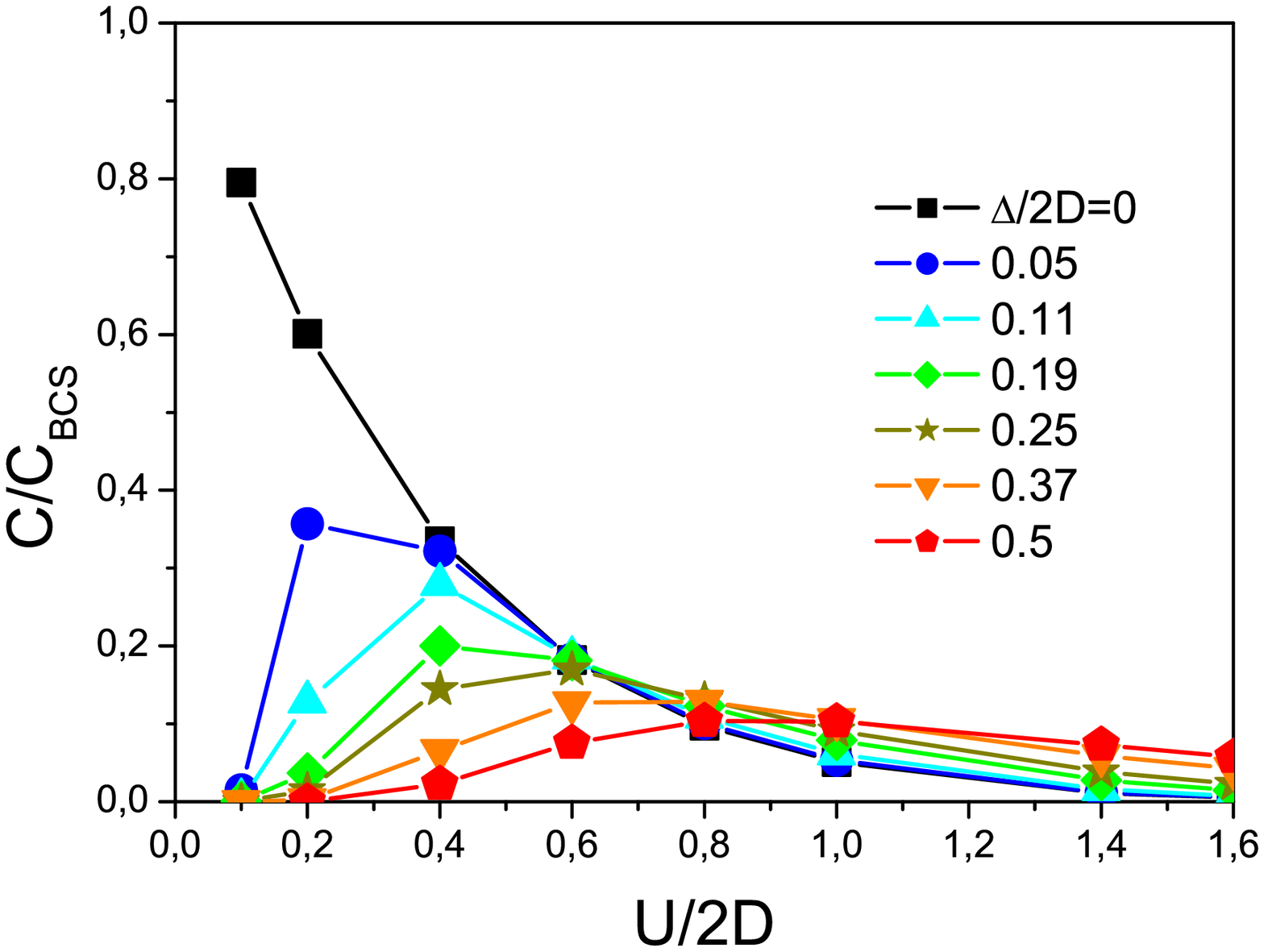}
\caption{Dependence of the coefficient $C$ on the strength of Hubbard attraction
for different levels of disorder ($a$ -- lattice parameter).
(a) --- all energy units are normalized by the width of the ``bare''
conduction band $2D$.
(b) --- all energy units normalized by effective band width $2D_{eff}$.
(c) --- coefficient $C$ normalized by its value $C_{BCS}$ in the weak coupling
limit and in the absence of disorder.
}
\label{fig5}
\end{figure}

In Fig. \ref{fig5} we show the dependencies of coefficient $C$ on the strength
of Hubbard attraction for different disorder levels. It is seen that $C$ drops
fast with the growth of the coupling constant. Especially fast this drop is
in the weak coupling region (see insert in Fig. \ref{fig5}(a)).
Being essentially a two -- particle characteristic coefficient $C$ does not
demonstrate universal dependencies on disorder, similar to $\alpha$ and $B$
coefficients, as is clearly seen from Fig. \ref{fig5} (b). Fig. \ref{fig5} (c)
shows the coupling strength dependence of $C$ normalized by its BCS value
(\ref{C_BCS}) in the absence of disorder.

\begin{figure}
\includegraphics[clip=true,width=0.7\textwidth]{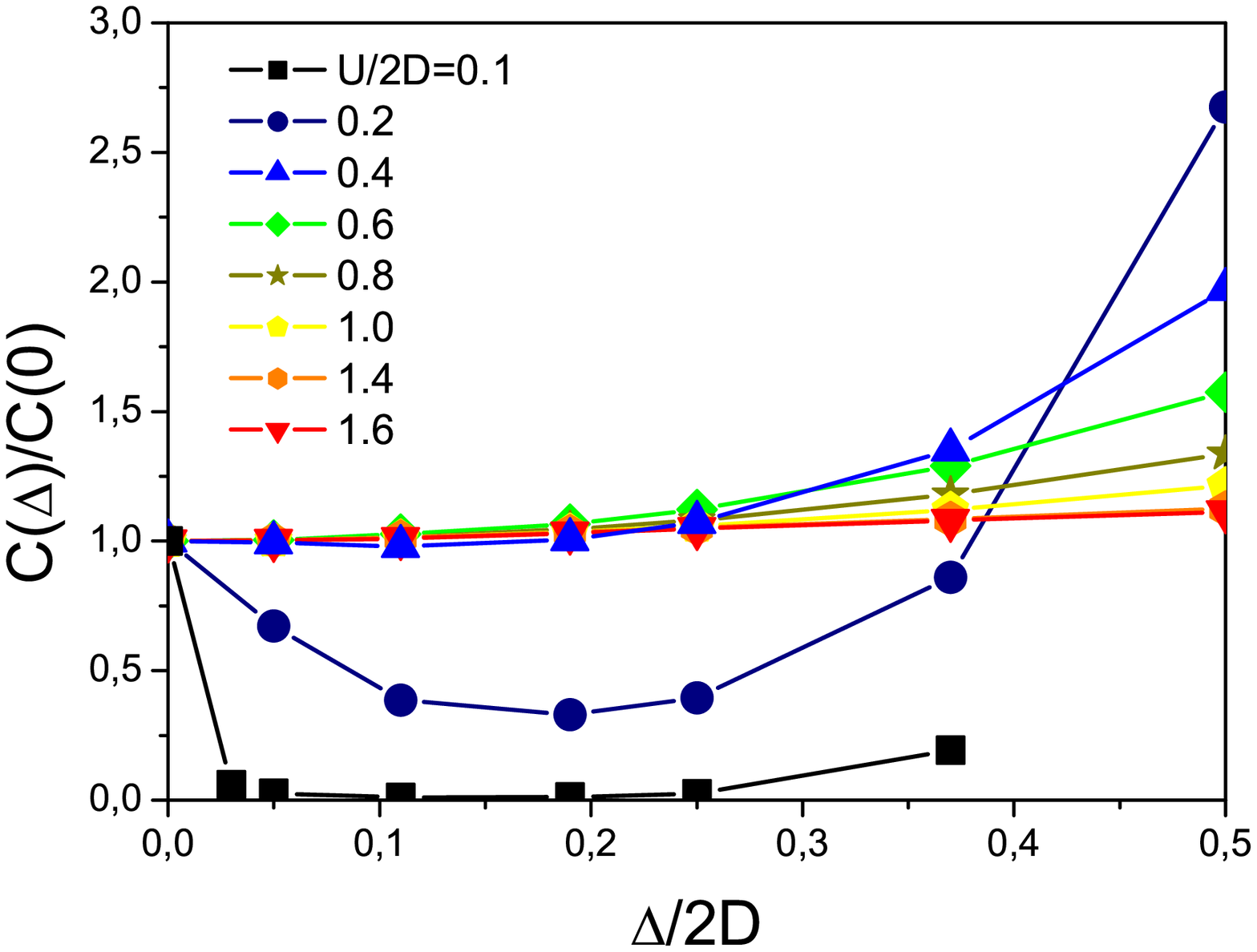}
\caption{Dependence of coefficient $C$, normalized by its value in the absence
of disorder, on disorder levels for different values of Hubbard attraction $U$.}
\label{fig6}
\end{figure}

In Fig. \ref{fig6} we show the dependencies of $C$ on disorder for different
values of coupling strength $U/2D$. In the weak coupling limit ($U/2D=0.1$) we
observe fast enough drop of $C$ with the growth of disorder in the region of
weak enough disorder scattering. However, in the region of strong enough
disorder we can observe even the growth of $C$ with disorder, related mainly to
noticeable band widening at high disorder levels and respective drop in the
effective coupling $U/2D_{eff}$. For intermediate couplings ($U/2D=0.4 - 0.6$)
coefficient $C$ only demonstrates some weak growth with disorder.
In BEC limit ($U/2D>1$) coefficient $C$ is practically independent of
disorder.

\begin{figure}
\includegraphics[clip=true,width=0.7\textwidth]{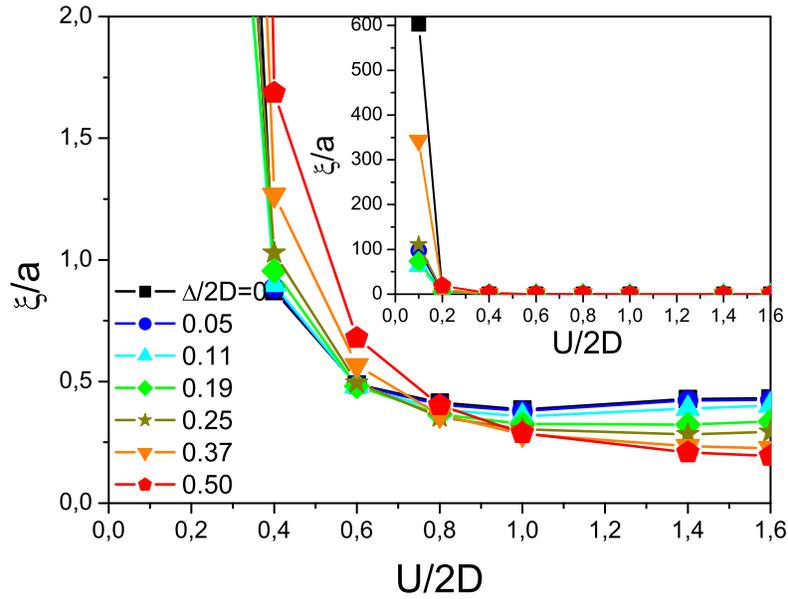}
\caption{Dependence of coherence length, normalized by lattice parameter $a$,
on Hubbard attraction $U$ for different disorder levels. Insert: fast growth of
coherence length in weak coupling BCS limit.}
\label{fig7}
\end{figure}

Let us now discuss the physical characteristics. Dependence of coherence length on
the strength of Hubbard attraction is shown in Fig. \ref{fig7}. We can see that
in the weak coupling region (cf. insert in Fig.\ref{fig7}) the coherence length
drops fast with the growth of $U$ at any disorder level, reaching the values
of the order of lattice spacing $a$ at the intermediate couplings
$U/2D \sim 0.4-0.6$. The further growth of the coupling strength leads only to
small changes of coherence length.

\begin{figure}
\includegraphics[clip=true,width=0.7\textwidth]{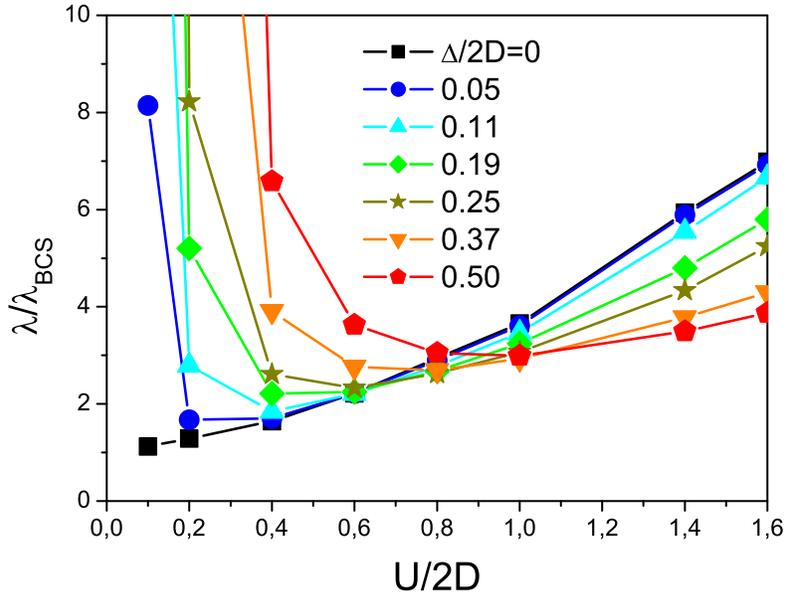}
\caption{Dependence of penetration depth, normalized by its BCS value in the
absence of disorder, on Hubbard attraction $U$ for different disorder levels.}
\label{fig8}
\end{figure}

In Fig. \ref{fig8} we show the dependence of penetration depth, normalized by its
BCS value in the absence of disorder (\ref{lambda_BCS}), on Hubbard attraction
$U$ for different levels of disorder.  In the absence of disorder scattering
penetration depth grows with coupling. Disorder in BCS weak coupling limit
leads to fast growth of penetration depth (for ``dirty'' BCS superconductors
$\lambda\sim l^{-1/2}$, where $l$ is the mean free path). In BEC strong coupling
region disorder only slightly diminishes the penetration depth (cf. Fig. \ref{fig11}(a)).

\begin{figure}
\includegraphics[clip=true,width=0.7\textwidth]{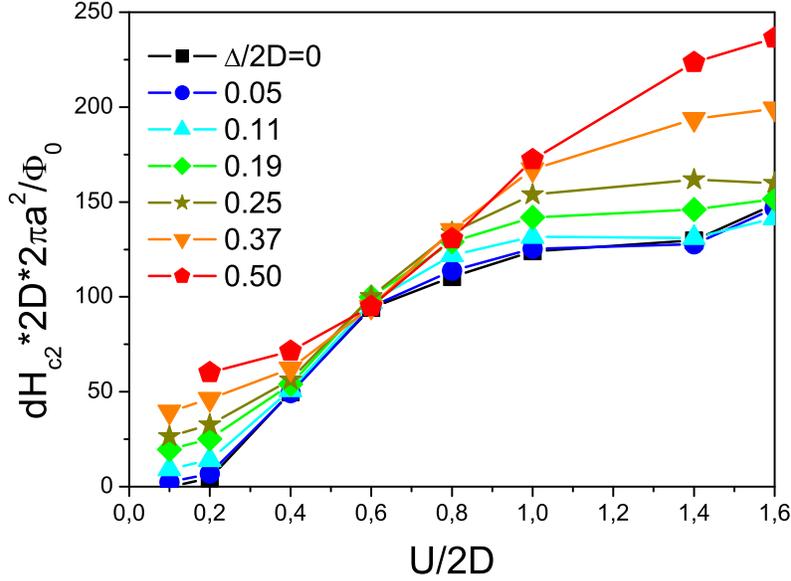}
\caption{Dependence of the upper critical magnetic field slope on Hubbard
attraction $U$ for different disorder levels.}
\label{fig9}
\end{figure}

Dependence of the slope of the upper critical filed 
$dH_{c2}\equiv (dH_{c2}/dT)_{T=T_c}$
on Hubbard attraction for
different disorder levels is shown in Fig. \ref{fig9}. For any value of disorder
scattering the slope of the upper critical field  grows with coupling. However,
in the limit of weak disorder we observe the fast growth of the slope with
$U$ in the limit of weak enough attraction, while in the strong coupling limit
the slope is weakly dependent on $U/2D$.

\begin{figure}
\includegraphics[clip=true,width=0.48\textwidth]{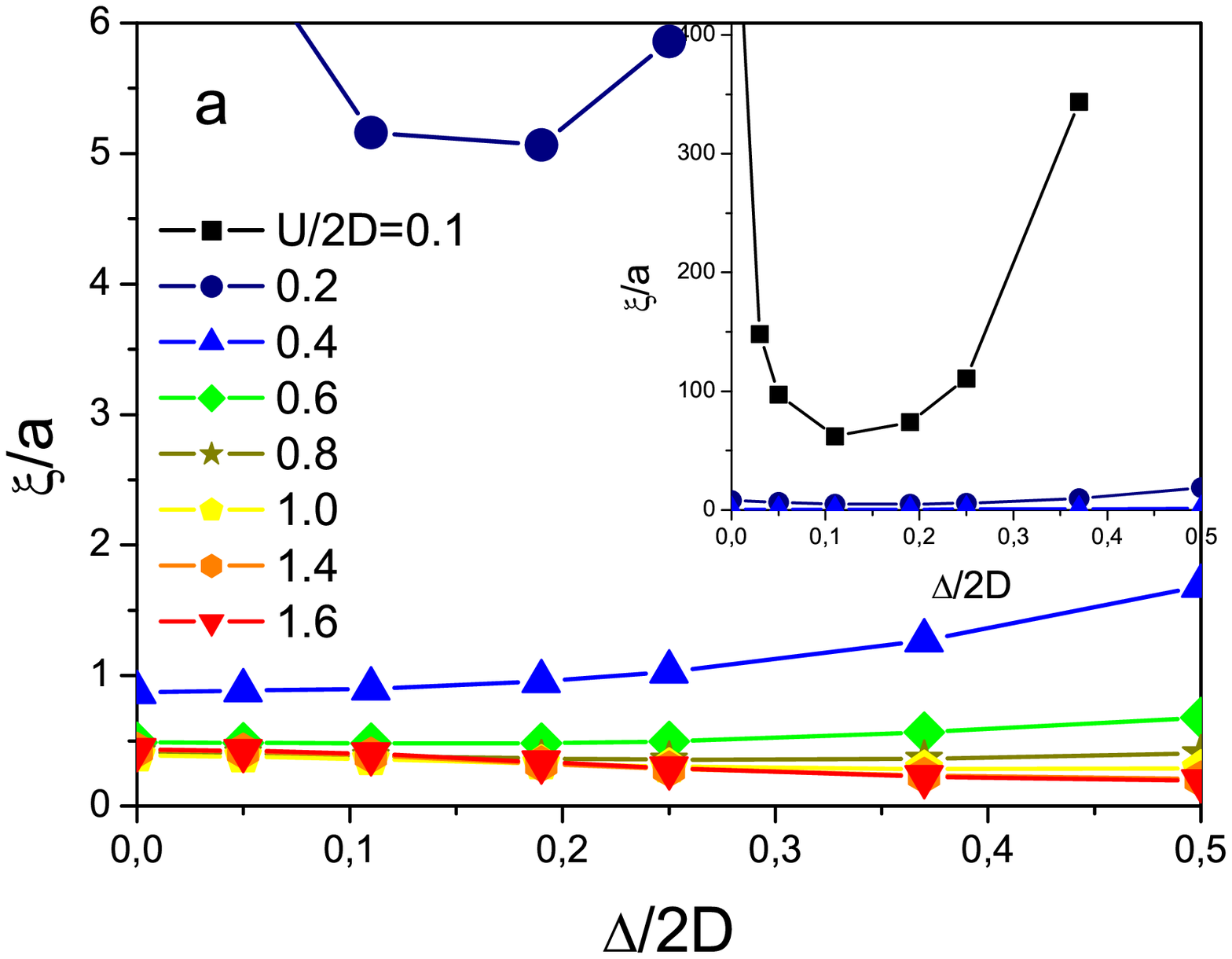}
\includegraphics[clip=true,width=0.48\textwidth]{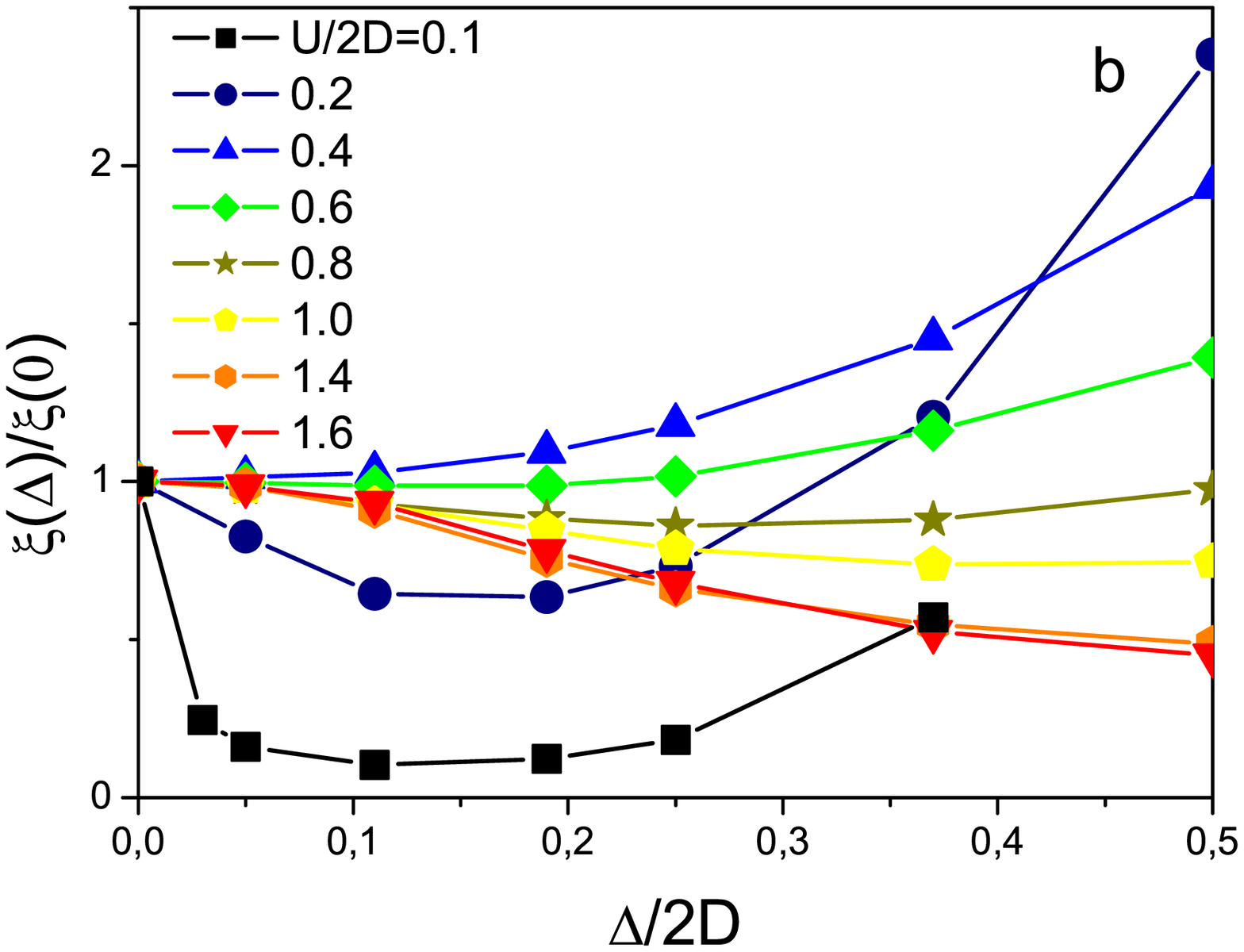}
\caption{Dependence of coherence length on disorder for different values of Hubbard
attraction.
(a) --- coherence length normalized by lattice parameter $a$.
Insert: coherence length dependence on disorder in the limit of weak coupling.
(b) --- coherence length normalized by its value in the absence of disorder.}
\label{fig10}
\end{figure}

In Fig. \ref{fig10} we show the dependence of coherence length $\xi$ on disorder
for different values of coupling. In BCS weak coupling limit and for weak
enough disorder we observe the standard ``dirty'' superconductors dependence
$\xi \sim l^{1/2}$, i.e. the coherence length drops with the growth of disorder
(cf. insert in Fig. \ref{fig10}(a)). However, for strong enough disorder the
coherence length starts to grow with disorder (cf. insert in Fig. \ref{fig10}(a)
and  Fig. \ref{fig10}(b)), which is mainly related to the noticeable widening
of the initial band by disorder and appropriate drop of $U/2D_{eff}$.
With further growth of the coupling strength $U/2D \geq 0.4-0.6$ the
coherence length $\xi$ becomes of the order of the lattice parameter and is
almost independent of disorder. In particular, in strong coupling BEC limit for
$U/2D=1.4, 1.6$ the growth of disorder to very large values ($\Delta/2D=0.5$)
leads to the drop of coherence length by the factor of two (cf. Fig. \ref{fig10}(b)).

\begin{figure}
\includegraphics[clip=true,width=0.48\textwidth]{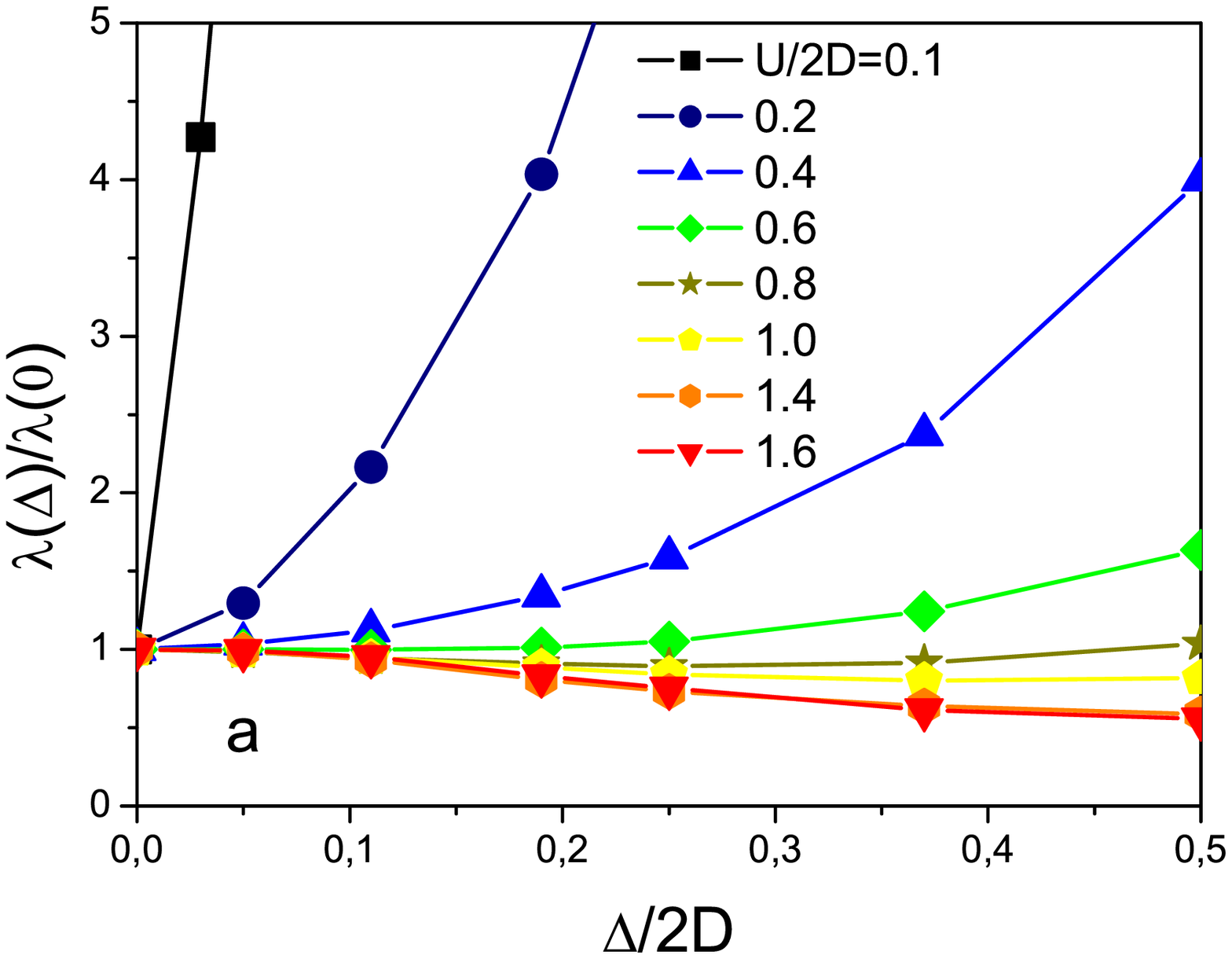}
\includegraphics[clip=true,width=0.48\textwidth]{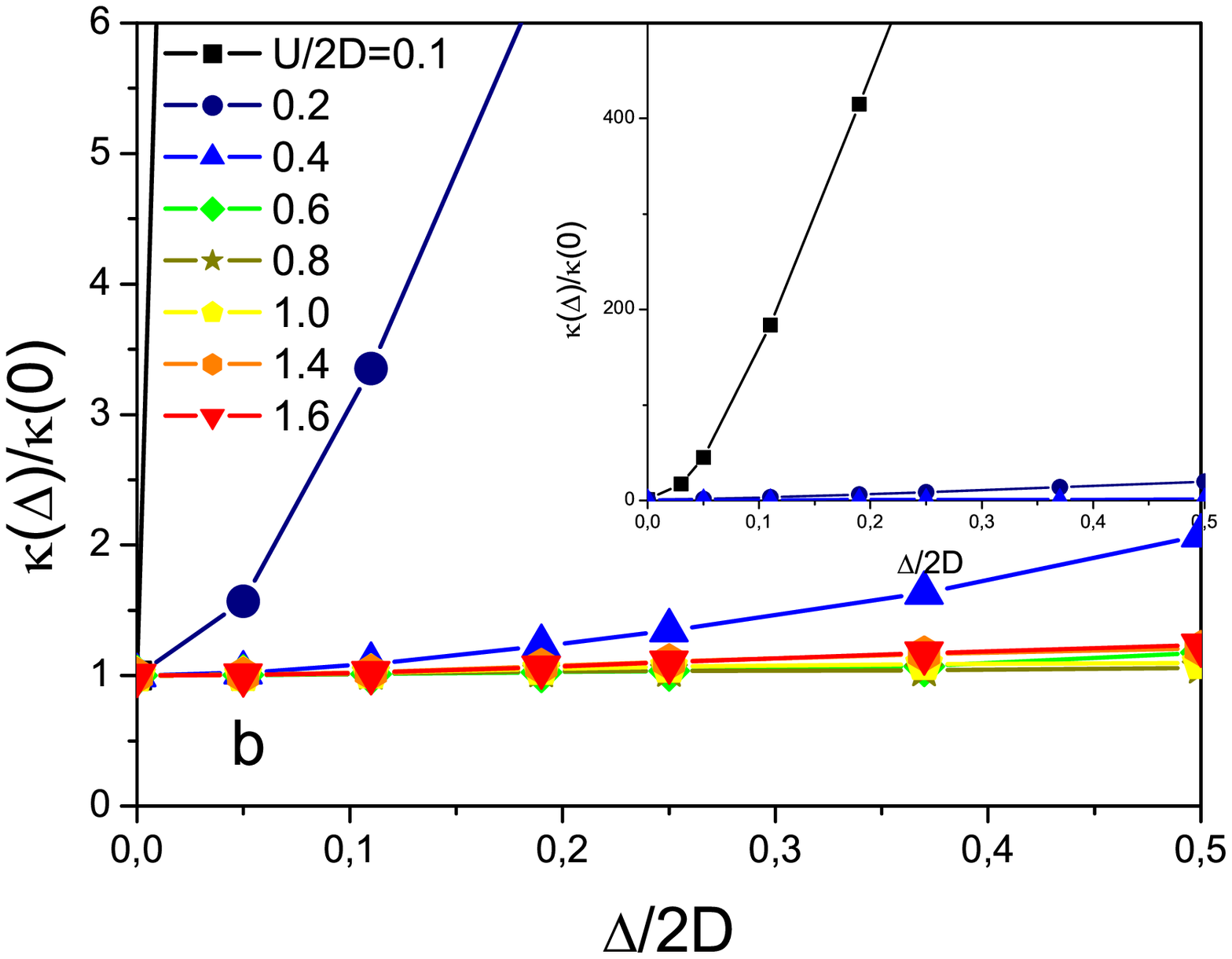}
\caption{Dependence of penetration depth (a) and Ginzburg -- Landau parameter (b)
on disorder for different values of Hubbard attraction. Insert: the growth of
GL -- parameter with disorder in the weak coupling limit.}
\label{fig11}
\end{figure}

Dependence of penetration depth on disorder for different values of Hubbard
attraction is shown in Fig. \ref{fig11}(a). In the limit of weak coupling in
accordance with the theory of ``dirty'' superconductors disorder leads to the
growth of penetration depth $\lambda \sim l^{-1/2}$. With the increase of 
the coupling strength this growth of penetration depth with disorder slows down 
and in the limit of very strong coupling $U/2D=1.4, 1.6$ penetration depth even 
slightly diminishes with the growth of disorder. In Fig. \ref{fig11}(b) we show 
the disorder dependence of dimensionless Ginzburg -- Landau parameter
$\kappa = \lambda / \xi$. We can see that in the weak coupling limit GL --
parameter grows fast with disorder (cf. insert in Fig. \ref{fig11}(b)) in
accordance with the theory of ``dirty'' superconductors, where $\kappa \sim l^{-1}$.
With the increase of the coupling the growth of GL -- parameter with disorder slows
down and in the limit of strong coupling $U/2D>1$ parameter $\kappa$ is
practically independent of disorder.

\begin{figure}
\includegraphics[clip=true,width=0.48\textwidth]{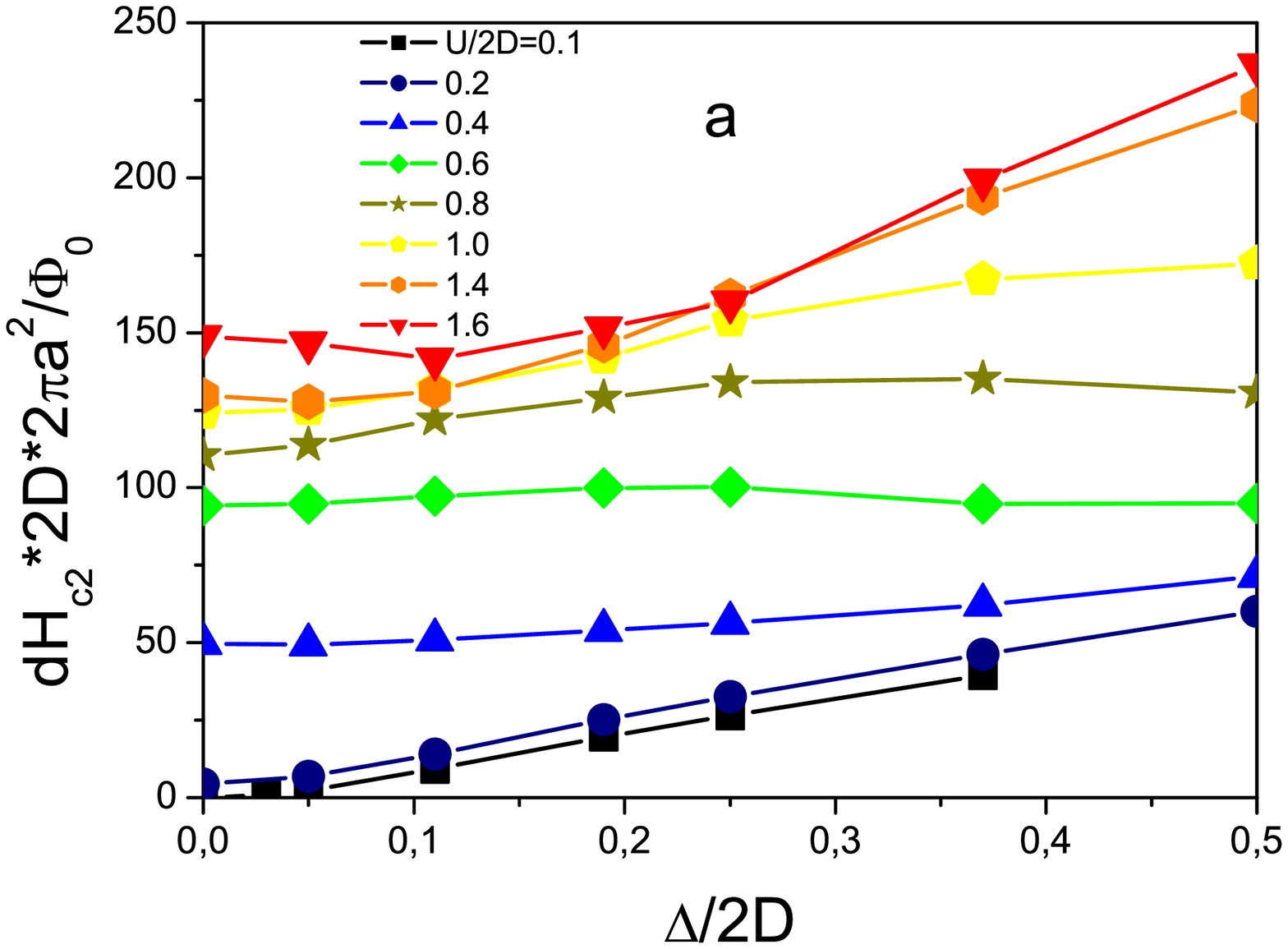}
\includegraphics[clip=true,width=0.48\textwidth]{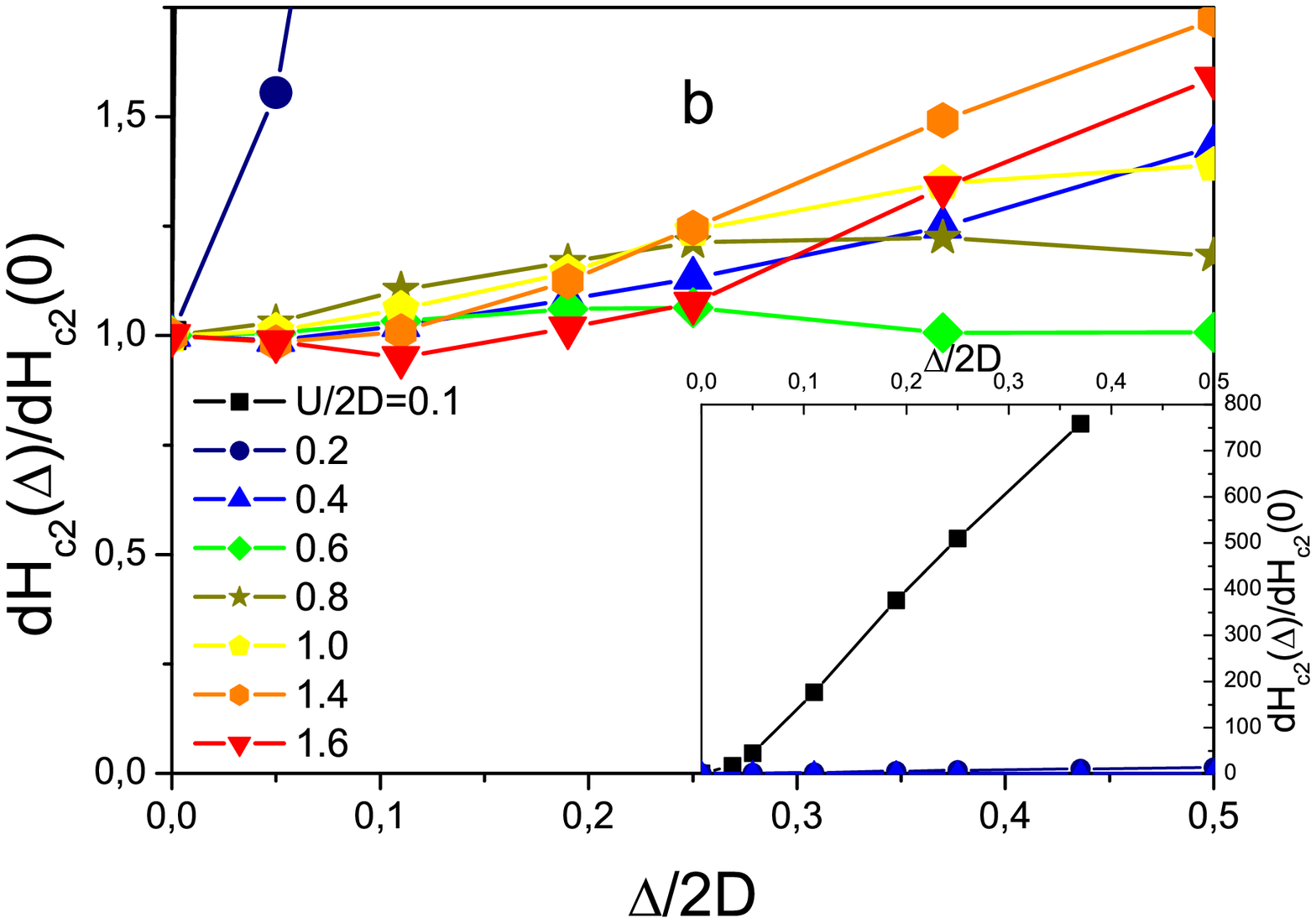}
\caption{Dependence of the slope of the upper critical field (a) and the
slope normalized by its value in the absence of disorder (b) on the level of
disorder for different values of Hubbard attraction. Insert:
the growth of the slope with disorder in the weak coupling limit.}
\label{fig12}
\end{figure}

In Fig. \ref{fig12} we show the dependence of the slope of the upper critical
magnetic field on disorder. In the weak coupling limit we again observe the typical
``dirty'' superconductor behavior --- the slope grows with disorder
(cf. Fig. \ref{fig12}(a) and the insert in Fig. \ref{fig12}(b)). For the
intermediate coupling region ($U/2D=0.4 - 0.8$) the slope of the upper critical
field is practically independent of disorder. In the limit of very strong
coupling  at small disorder the slope of the upper critical field can even
slightly diminish, but in the limit of strong disorder the slope grows with
the growth of disorder scattering.

\section{Conclusion}

In the framework of DMFT+$\Sigma$ generalization of dynamic mean field theory we
have studied the effects of disorder on the coefficients of Ginzburg -- Landau
expansion and the related physical characteristics of superconductors close to
$T_c$ in attractive Hubbard model. To study the GL -- coefficients we have used
the combination of  DMFT+$\Sigma$ approach and Nozieres -- Schmitt-Rink
approximation. Calculations were performed for the wide range of the values of
attractive potential $U$, from the weak coupling region ($U/2D_{eff}\ll 1$),
where instability of the normal phase and superconductivity are well described
by BCS model, up to the strong coupling limit ($U/2D_{eff}>1$), where the
superconducting transition is related to the Bose -- Einstein condensation of
compact Cooper pairs.

The growth of the coupling strength $U$ leads to fast drop of all GL -- coefficients.
Coherence length $\xi$ drops fast with the growth of the coupling strength and
for $U/2D \sim 0.4$ becomes of the order of the lattice parameter and only
slightly changes with the further growth of the coupling.  Penetration depth
in ``clean'' superconductors grows with $U$, while in ``dirty'' case it
drops in the weak coupling region and grows in BEC limit, passing through the
minimum in the intermediate (crossover) region of $U/2D\sim 0.4-0.8$. The slope
of the upper critical magnetic field grows with $U$. Specific heat discontinuity
grows with Hubbard attraction $U$ in the weak coupling region and diminishes
in the strong coupling region, passing through the maximum at $U/2D_{eff}\approx 0.55$.

Disorder influence on the critical temperature $T_c$, GL -- coefficients $A$ and
$B$ and specific heat discontinuity is universal --- their change is related only
to conduction band widening by disorder scattering, i.e. to the replacement
$D \to D_{eff}$. Thus, both in BCS -- BEC crossover region and in the strong
coupling limit both critical temperature and GL -- coefficients $A$ and $B$ obey
the generalized Anderson theorem --- all the influence of disorder reduces to
disorder change of the density of states.

GL -- coefficient $C$ was studied here in the ``ladder'' approximation for
disorder scattering. Disorder influence upon $C$ is not universal and is not
related purely to the conduction band widening by disorder. In the limit of
weak coupling $U/2D_{eff}\ll 1$ the behavior of $C$ and the related physical
characteristics are well described by the usual theory of ``dirty'' superconductors.
Both $C$ and coherence length drops fast with the growth of disorder, while the
penetration depth and the slope of the upper critical magnetic field grow with
disorder. In the region of BCS -- BEC crossover and in the BEC limit
the coefficient $C$ and all physical characteristics are only weakly dependent
on disorder. In particular, in BEC limit both the coherence length and
penetration depth are only slightly suppressed with the growth of disorder, so
that the GL -- parameter $\kappa$ is practically independent of disorder.

This work was supported by RSF grant 14-12-00502.


\newpage

\end{document}